# A New Proposed Dynamic Quantum with Re-Adjusted Round Robin Scheduling Algorithm and Its Performance Analysis

H.S.Behera
Professor
Veer Surendra Sai University of Technology,Burla
Sambalpur,India

R. Mohanty
Professor
Veer Surendra Sai University of Technology,Burla
Sambalpur,India

Debashree Nayak
Research Associate
Veer Surendra Sai University of Technology,Burla
Sambalpur,India

## ABSTRACT
Scheduling is the central concept used frequently in Operating System. It helps in choosing the processes for execution. Round Robin (RR) is one of the most widely used CPU scheduling algorithm. But, its performance degrades with respect to context switching, which is an overhead and it occurs during each scheduling. Overall performance of the system depends on choice of an optimal time quantum, so that context switching can be reduced. In this paper, we have proposed a new variant of RR scheduling algorithm, known as Dynamic Quantum with Re-adjusted Round Robin (DQRRR) algorithm. We have experimentally shown that performance of DQRRR is better than RR by reducing number of context switching, average waiting time and average turnaround time.

## 1. INTRODUCTION
Operating system is an interface between end user and system hardware, so that the user can handle the system in a convenient manner. In a single user environment, there was no need to choose any task because task execution continues one after another, but in multitasking environment, it becomes necessary for the processor to choose a task from the ready queue. Operating system follows a predefined procedure for selecting process among number of processes from the ready queue, known as Scheduling. Scheduler selects the ready processes from memory and allocates resource/CPU as per their requirement. Whenever one process waits for some other resource, scheduler selects next process and allocates CPU to it. This process continues till the system request for termination of execution and then the last CPU burst ends up with it.

### 1.1 Scheduling Algorithms
In the First-Come-First-Serve (FCFS) algorithm, process that arrives first is immediately allocated to the CPU based on FIFO policy. In Shortest Job First (SJF) algorithm, process having shortest CPU burst time will execute first. If two processes having same burst time and arrive simultaneously, then FCFS procedure is applied. Priority scheduling algorithm, provides priority (internally or externally) to each process and selects the highest priority process from the ready queue. In case of Round Robin (RR) algorithm, time interval of one time quantum is given to each process present in the circular queue emphasizing on the fairness factor.

### 1.2 Motivation
In RR scheduling fairness is given to each process, i.e. processes get fair share of CPU because of given time slice. So, it is better than other scheduling algorithms. Number of context switching incase of RR Scheduling is n in one round only, i.e. high in comparison to other scheduling algorithms. It gives low turnaround time and average waiting time. RR scheduling uses static time quantum that gives large waiting time and turnaround time in case of variable burst time which degrades the overall performance. This factor motivates us to design an improved algorithm which can overcome the above limitation.

### 1.3 Related Work
SARR algorithm [1] is based on a new approach called dynamic-time-quantum, in which time quantum is repeatedly adjusted according to the burst time of the running processes. Mixed Scheduling (A New Scheduling Policy) [2], uses the job mix order for non preemptive scheduling FCFS and SJF. According to job mix order, from a list of N processes, the process which needs minimum CPU time is executed first and then the highest from the list and so on till the nth process. In Burst Round Robin (BRR) [3], a new weighting technique is introduced for CPU Schedulers. Here shorter jobs are given more time, so that processes having shorter jobs are cleared from the ready queue in a short time span.

### 1.4 Our Contribution
In this paper, the principal objective is to reduce context switching occur in RR scheduling. For that purpose, we have developed a method that drastically reduces context switching.

### 1.5 Organization of the Paper
This paper presents the method for reducing context switch, average waiting time and average turnaround time using random sorting and dynamic time quantum. Section 2 discusses background preliminaries. Section 3 presents the proposed approach. Section 4 shows experimental analysis. In Section 5 conclusion and future work towards our method is given.

## 2. BACKGROUND WORK
### 2.1 Terminologies
Burst time ($b_t$) is the time needed by the process to hold the control of CPU. Time Quantum ($q_t$) is a particular slice of time given to each process to have CPU for that time period only. Average Waiting Time ($a_{wt}$) is the time gap between arrival of one process and its response by the CPU. To achieve good result, $a_{wt}$ should be less. Average Turnaround Time ($a_{tat}$) is the time gap between the instant of process arrival and the instant of its completion. For getting good result, it should be less. Context Switch (CS) is the number of time CPU switches from one process to another. For better performance of the algorithm, it should be less.

### 2.2 RR Scheduling Algorithm
RR Scheduling Algorithm is the simplest and widely used algorithm as it gives fairness to each process. Newly arrived processes are kept in the rear part of the queue. Scheduler chooses each process from front of the queue and allocates the CPU for one time quantum. The performance of RR algorithm depends heavily on the size of the time quantum [1]. For smaller time quantum, the context switching is more and for larger time quantum, response time is more. Overall performance of RR may decrease for weak time quantum selection. Therefore, choice of an optimal time quantum is necessary.

## 3. PROPOSED APPROACH
In proposed approach, we have to arrange the processes in ascending order according to their burst time present in the ready queue. Then time quantum is calculated. For finding an optimal time quantum, median method is followed. The median can be found out using the following formulae [1].

$$\text{Median } \bar{x} = \begin{cases} Y_{\frac{n+1}{2}} & \text{if n is odd} \\ 1/2[Y_{n/2} + Y_{1+\frac{n}{2}}] & \text{if n is even} \end{cases}$$

Where, $\bar{x}$ = median

y = number located in the middle
   of a group of numbers arranged
   in ascending order

n = number of processes

Here, the time quantum is assigned to the processes. This time quantum is recalculated taking the remaining burst time in account after each cycle. In the next step we have to rearrange the sorted processes, i.e. among n processes, the process which needs minimum CPU burst time will be replaced as the first process and then the process with highest CPU burst time from the queue, will be replaced as the second process and so on.

### 3.1 Proposed Algorithm
1. I/P: Process ($P_n$), Burst Time ($b_t$), Arrival Time ($a_t$), ready queue.
   O/P: Context Switch (cs), Average Waiting Time ($a_{wt}$), Average Turnaround time ($a_{tat}$).
2. Initialize: ready queue=0, cs=0, $a_{wt}$=0, $a_{tat}$=0.
3. while (ready queue!=NULL)
       sort the processes in ascending order according to their $b_t$
       //find median
           $q_t$=median
4. //sort the processes
   for each process i=1 to n
       do {
       if (i%2==0)
           put the list amount $b_t$ in ready queue
       else
           put the highest amount $b_t$ in the ready queue
       } end for
5. // assign qt to each process
       for each process i=1 to n
           P[i] → $q_t$
6. // if a new process arrives
       update the counter n and go to step 2.
       end while
7. $a_{wt}$, $a_{tat}$ and cs are calculated
8. stop & exit

### 3.2 Illustration
To demonstrate the above algorithm we have considered the following example. Arrival time is considered to be zero for the given processes P1, P2, P3, P4 and corresponding burst times are 21, 105, 12, 55 respectively. In first step the processes in the ready queue are sorted in ascending order. Then the time quantum is calculated in the second step. Here $q_t$ = 38. In third step sorted processes are rearranged as described in the 3$^{rd}$ section, i.e. P3 with $b_t$=12, P2 having $b_t$= 105, P1 with $b_t$=21 and P4 with $b_t$= 55. After assigning $q_t$ to each process the remaining burst time of all process are P3=0, P2=67, P1=0 and P4=17. When a process completes its execution, it is deleted from ready queue automatically. Further the next time quantum is calculated from remaining burst times as per the 3$^{rd}$ step in the algorithm. Here $q_t$=42. Then the remaining burst times are P2=25 and P4 =0. According to the algorithm the next $q_t$ will be 25 and in the last step the process P2 will complete its execution and will be deleted from the ready queue.

## 4. EXPERIMENTAL ALANALYSIS
### 4.1 Assumptions
Our experiments are performed in a uni-processor environment and the processes taken are CPU bound processes only. Here we have taken n processes, i.e. P1, P2… Pn and all these processes are independent from each other. For all the processes, corresponding burst time ($b_t$) and arrival time ($a_t$) are known before submitting the processes to the processor.

### 4.2 Experimental Frame Work
The input parameters taken are as follows. $P_n$ is the number of processes. $a_t$, $b_t$, $q_t$ are the arrival time, burst time and quantum time respectively. The output parameters are context switch(CS), average waiting time($a_{wt}$) and average turnaround time($a_{tat}$). We have taken two cases, i.e. case 1 is for processes with zero arrival time (here each process arrive at same time) and case 2 is for processes without zero arrival time (here processes are arrived at different time). Under these two cases we have performed three different experiments taking three different types of data sets (data sets in increasing order, decreasing order and random order).

## 4.3 Results Obtained

This algorithm can work effectively with large number of data. In each case we have compared our proposed algorithm's results with Round Robin scheduling algorithm's result. For RR Scheduling Algorithm we have taken 25 as the fixed time quantum.

## Case 1: With Zero Arrival Time

### Increasing Order

We consider five processes P1, P2, P3, P4 and P5 arriving at time 0 with burst time 30, 42, 50, 85, 97 respectively shown in Table 4.1. Table 4.2 shows the comparing result of RR algorithm and our proposed algorithm (DQRRR).

**Table 4.1. Data in Increasing Order**

| No. of process | $a_t$ | $b_t$ |
|---|---|---|
| P1 | 0 | 30 |
| P2 | 0 | 42 |
| P3 | 0 | 50 |
| P4 | 0 | 85 |
| P5 | 0 | 97 |

**Table 4.2. Comparison between RR and DQRRR**

| algorithms | RR | DQRRR |
|---|---|---|
| $q_t$ | 25 | 50,41,6 |
| CS | 13 | 7 |
| $a_{wt}$ | 146.2 | 134.4 |
| $a_{tat}$ | 207 | 195.2 |

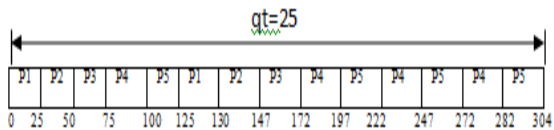

**Fig.4.2: Gantt chart for RR in Table 4.2**

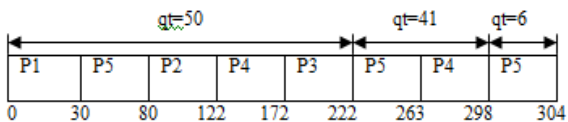

**Fig.4.2: Gantt chart for DQRRR in Table 4.2**

### Decreasing Order

We consider five processes P1, P2, P3, P4 and P5 arriving at time 0 with burst time 105, 90, 60, 45, 35 respectively shown in Table 4.3. Table 4.4 shows the comparing result of RR algorithm and our proposed algorithm (DQRRR).

**Table 4.3. Data in Decreasing Order**

| No. of process | $a_t$ | $b_t$ |
|---|---|---|
| P1 | 0 | 105 |
| P2 | 0 | 90 |
| P3 | 0 | 60 |
| P4 | 0 | 45 |
| P5 | 0 | 35 |

**Table 4.4. Comparison between RR and DQRRR**

| algorithms | RR | DQRRR |
|---|---|---|
| $q_t$ | 25 | 60,37,8 |
| CS | 15 | 7 |
| $a_{wt}$ | 214 | 152.4 |
| $a_{tat}$ | 281 | 219.4 |

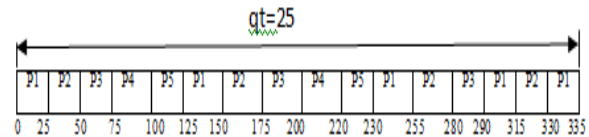

**Fig.4.3: Gantt chart for RR in Table 4.4**

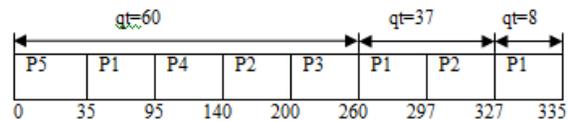

**Fig.4.4: Gantt chart for DQRRR in Table 4.4**

### Random Order

We consider five processes P1, P2, P3, P4 and P5 arriving at time 0 with burst time 92, 70,35,40,80 respectively shown in Table 4.5. Table 4.6 shows the comparing result of RR algorithm and our proposed algorithm (DQRRR).

**Table 4.5. Data in Random Order**

| No. of process | $a_t$ | $b_t$ |
|---|---|---|
| P1 | 0 | 92 |
| P2 | 0 | 70 |
| P3 | 0 | 35 |
| P4 | 0 | 40 |
| P5 | 0 | 80 |

have taken two cases, i.e. case 1 is for processes with zero arrival time (here each process arrive at same time) and case 2 is for processes without zero arrival time (here processes are arrived at different time). Under these two cases we have performed three different experiments taking three different types of data sets (data sets in increasing order, decreasing order and random order).

## 4.3 Results Obtained

This algorithm can work effectively with large number of data. In each case we have compared our proposed algorithm's results with Round Robin scheduling algorithm's result. For RR Scheduling Algorithm we have taken 25 as the fixed time quantum.

## Case 1: With Zero Arrival Time

### Increasing Order

We consider five processes P1, P2, P3, P4 and P5 arriving at time 0 with burst time 30, 42, 50, 85, 97 respectively shown in Table 4.1. Table 4.2 shows the comparing result of RR algorithm and our proposed algorithm (DQRRR).

**Table 4.1. Data in Increasing Order**

| No. of process | $a_t$ | $b_t$ |
|---|---|---|
| P1 | 0 | 30 |
| P2 | 0 | 42 |
| P3 | 0 | 50 |
| P4 | 0 | 85 |
| P5 | 0 | 97 |

**Table 4.2. Comparison between RR and DQRRR**

| algorithms | RR | DQRRR |
|---|---|---|
| $q_t$ | 25 | 50,41,6 |
| CS | 13 | 7 |
| $a_{wt}$ | 146.2 | 134.4 |
| $a_{tat}$ | 207 | 195.2 |

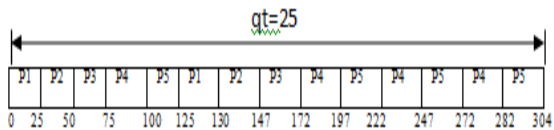

**Fig.4.2: Gantt chart for RR in Table 4.2**

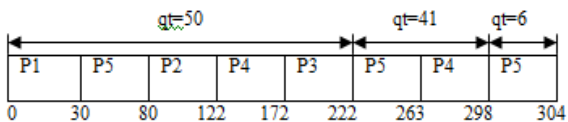

**Fig.4.2: Gantt chart for DQRRR in Table 4.2**

### Decreasing Order

We consider five processes P1, P2, P3, P4 and P5 arriving at time 0 with burst time 105, 90, 60, 45, 35 respectively shown in Table 4.3. Table 4.4 shows the comparing result of RR algorithm and our proposed algorithm (DQRRR).

**Table 4.3. Data in Decreasing Order**

| No. of process | $a_t$ | $b_t$ |
|---|---|---|
| P1 | 0 | 105 |
| P2 | 0 | 90 |
| P3 | 0 | 60 |
| P4 | 0 | 45 |
| P5 | 0 | 35 |

**Table 4.4. Comparison between RR and DQRRR**

| algorithms | RR | DQRRR |
|---|---|---|
| $q_t$ | 25 | 60,37,8 |
| CS | 15 | 7 |
| $a_{wt}$ | 214 | 152.4 |
| $a_{tat}$ | 281 | 219.4 |

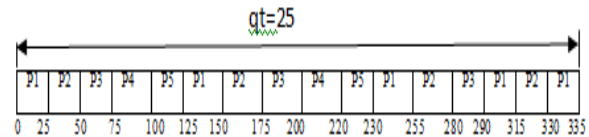

**Fig.4.3: Gantt chart for RR in Table 4.4**

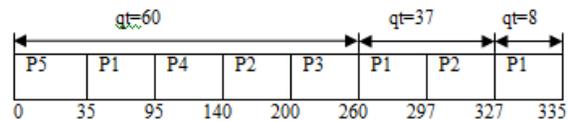

**Fig.4.4: Gantt chart for DQRRR in Table 4.4**

### Random Order

We consider five processes P1, P2, P3, P4 and P5 arriving at time 0 with burst time 92, 70,35,40,80 respectively shown in Table 4.5. Table 4.6 shows the comparing result of RR algorithm and our proposed algorithm (DQRRR).

**Table 4.5. Data in Random Order**

| No. of process | $a_t$ | $b_t$ |
|---|---|---|
| P1 | 0 | 92 |
| P2 | 0 | 70 |
| P3 | 0 | 35 |
| P4 | 0 | 40 |
| P5 | 0 | 80 |

**Table 4.6. Comparison between RR and DQRRR**

| algorithms | RR | DQRRR |
|---|---|---|
| $q_t$ | 25 | 80, 11,1 |
| CS | 14 | 7 |
| $a_{wt}$ | 173.4 | 150.2 |
| $a_{tat}$ | 256.8 | 215.6 |

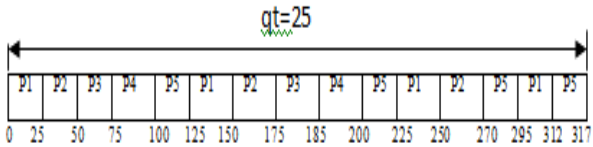

**Fig.4.5: Gantt chart for RR in Table 4.6**

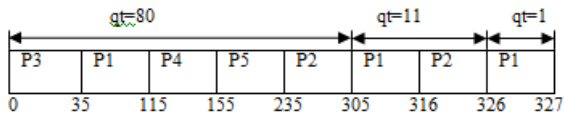

**Fig.4.6: Gantt chart for DQRRR in Table 4.6**

## Case 2: Without Zero Arrival Time

### Increasing Order

We consider five processes P1, P2, P3, P4 and P5 arriving at time 0,2,6,6,8 and burst time 28,35,50,82,110 respectively shown in Table 4.7. Table 4.8 shows the comparing result of RR algorithm and our proposed algorithm (DQRRR).

**Table 4.7. Data in Increasing Order**

| No.of process | $a_t$ | $b_t$ |
|---|---|---|
| P1 | 0 | 28 |
| P2 | 2 | 35 |
| P3 | 6 | 50 |
| P4 | 6 | 82 |
| P5 | 8 | 110 |

**Table 4.8. Comparison between RR and DQRRR**

| algorithms | RR | DQRRR |
|---|---|---|
| $q_t$ | 25 | 28,66,30,14 |
| CS | 14 | 7 |
| $a_{wt}$ | 139.8 | 112.2 |
| $a_{tat}$ | 199.4 | 173.2 |

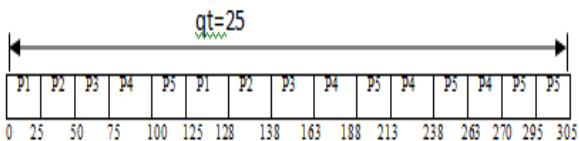

**Fig.4.7: Gantt chart for RR in Table 4.8**

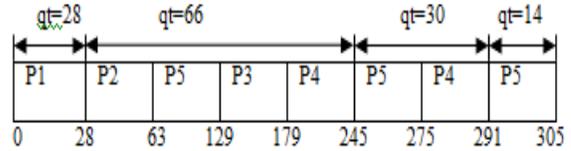

**Fig.4.8: Gantt chart for DQRRR in Table 4.8**

**Decreasing Order:**

We consider five processes P1, P2, P3, P4, P5 arriving at time 0,2,3,4,5 and burst Time80,72,65,50,43 respectively shown in Table 4.9. Table 4.10 shows the comparing result of RR algorithm and our proposed algorithm (DQRRR).

**Table 4.9. Data in Decreasing Order**

| No. of process | $a_t$ | $b_t$ |
|---|---|---|
| P1 | 0 | 80 |
| P2 | 2 | 72 |
| P3 | 3 | 65 |
| P4 | 4 | 50 |
| P5 | 5 | 43 |

**Table 4.10. Comparison between RR and DQRRR**

| Algorithms | RR | DQRRR |
|---|---|---|
| $q_t$ | 25 | 80,57,11,4 |
| CS | 13 | 7 |
| $a_{wt}$ | 216.8 | 147.8 |
| $a_{tat}$ | 280.2 | 209.8 |

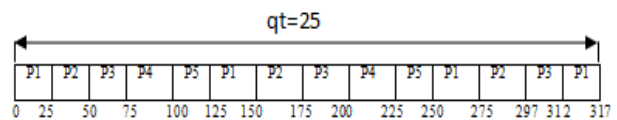

**Fig.4.9: Gantt chart for RR in Table 4.10**

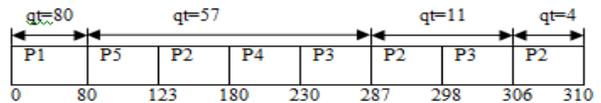

**Fig.4.10: Gantt chart for DQRRR in Table 4.10**

### Random Order

We consider five processes P1, P2, P3, P4 & P5 arriving at time 0, 1,2,5,7 and burst time 26,82,70,31,40 respectively shown in Table 4.11. Table 4.12 shows the comparing result of RR algorithm and our proposed algorithm (DQRRR)**.**

**Table 4.11.Data in Random Order**

| No. of process | $a_t$ | $b_t$ |
|---|---|---|
| P1 | 0 | 26 |
| P2 | 1 | 82 |
| P3 | 2 | 70 |
| P4 | 5 | 31 |
| P5 | 7 | 40 |

**Table 4.12.Comparison between RR and DQRRR**

| algorithms | RR | DQRRR |
|---|---|---|
| $q_t$ | 25 | 26,55,21,6 |
| CS | 12 | 7 |
| $a_{wt}$ | 149.4 | 95.6 |
| $a_{tat}$ | 199.2 | 145.4 |

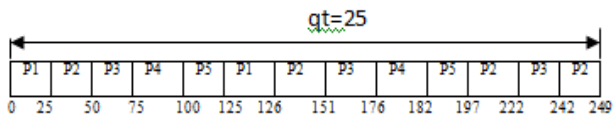

Fig.4.11: Gantt chart for RR in Table 4.12

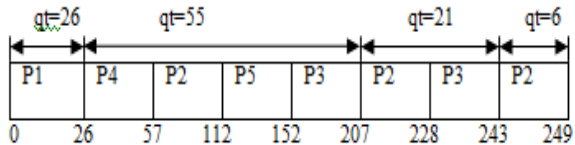

Fig.4.12: Gantt chart for DQRRR in Table 4.12

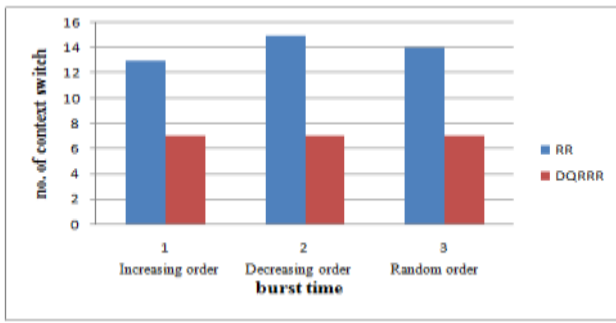

Fig.4.13: Context Switching(DQRRR vs. RR)

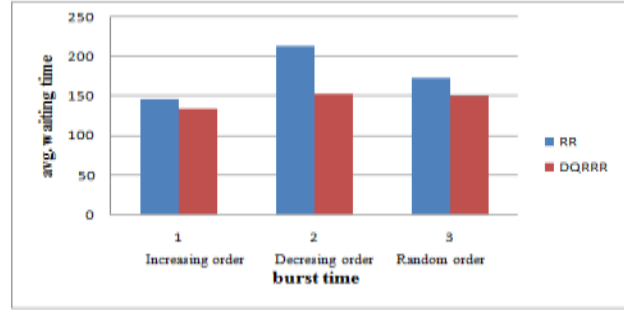

Fig.4.14: Average Waiting Time(DQRRR vs. RR)

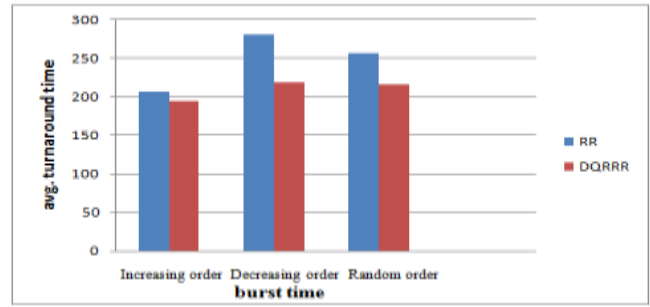

Fig.4.15: Average Turnaround Time(DQRRRvs.RR)

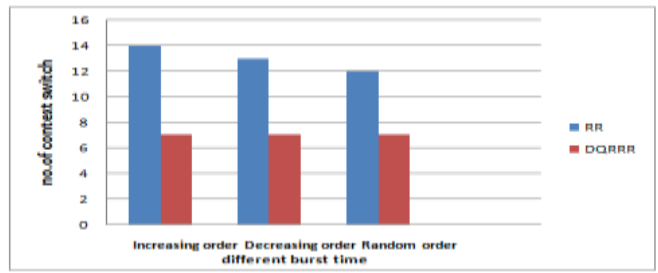

Fig.4.16: Context Switching(DQRRR vs. RR)

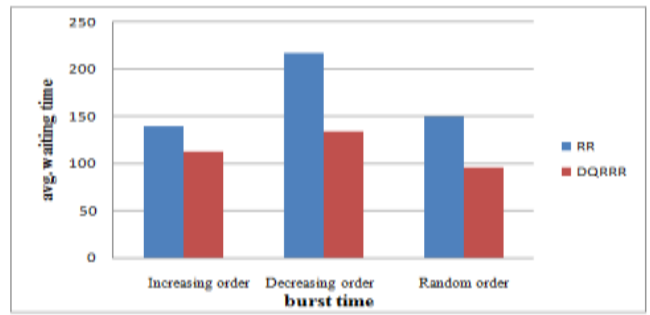

Fig.17: Average Waiting Time(DQRRR vs. RR)

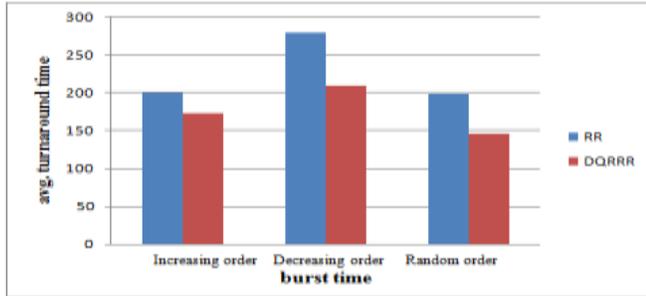

**Fig.18: Average Turnaround Time (DQRRR vs. RR)**

## 5. CONCLUSION AND FUTURE WORK

The proposed variant of RR algorithm drastically decreases context switching. The proposed algorithm performs better than RR scheduling algorithm with respect to average waiting time, turnaround time and context switching. Our proposed algorithm can be further investigated to be useful in providing more and more task-oriented results in future along with developing adaptive algorithms to fit the varying situations in today's multifaceted complex working of operating system.